# Switchable induced-transmission filters enabled by vanadium dioxide

Chenghao Wan[1,2†], David Woolf[3†], Colin M. Hessel[3], Jad Salman[1], Yuzhe Xiao[1], Chunhui Yao[1], Albert Wright[3], Joel M. Hensley[3], Mikhail A. Kats[1,2,4*]

[1]Department of Electrical and Engineering, University of Wisconsin-Madison, Madison, WI 53706, USA
[2]Department of Materials Science and Engineering, University of Wisconsin-Madison, Madison, WI 53706, USA
[3]Physical Sciences, Inc., Andover, MA
[4]Department of Physics, University of Wisconsin-Madison, Madison, WI 53706, USA
[†]*These authors contributed equally to this work*
*\*Corresponding author: mkats@wisc.edu*

**Abstract:** An induced-transmission filter (ITF) uses an ultrathin layer of metal positioned at an electric-field node within a dielectric thin-film bandpass filter to select one transmission band while suppressing other transmission bands that would have been present without the metal layer. Here, we introduce a switchable mid-infrared ITF where the metal film can be "switched on and off", enabling the modulation of the filter response from single-band to multiband. The switching is enabled by a deeply subwavelength film of vanadium dioxide ($VO_2$), which undergoes a reversible insulator-to-metal phase transition. We designed and experimentally demonstrated an ITF that can switch between two states: one broad passband across the long-wave infrared (LWIR, 8 – 12 μm) and one narrow passband at ~8.8 μm. Our work generalizes the ITF—previously a niche type of bandpass filter—into a new class of tunable devices. Furthermore, our unique fabrication process—which begins with thin-film $VO_2$ on a suspended membrane—enables the integration of $VO_2$ into any thin-film assembly that is compatible with physical vapor deposition (PVD) processes, and is thus a new platform for realizing tunable thin-film filters.

## Introduction

Various approaches have been employed to make optical filters tunable with respect to parameters such as wavelength, bandwidth, and transmission magnitude. Examples such as mechanically driven filter wheels[1], tunable fiber Bragg gratings[2,3], tunable Fabry-Pérot (F-P) resonators[4,5], acousto-optic tunable filters[6,7], and liquid-crystal tunable filters[8,9] have resulted in revolutionary developments in applications ranging from optical communications[10] to detection and imaging[11,12]. More recently, efforts to create fast, compact, and tunable filters have focused on incorporating active media into either multilayer[13,14,15] or metasurface-type[16,17,18] structures. Each of these approaches has different advantages; for example, metasurface-based designs are much thinner and require fewer discrete fabrication steps, whereas multilayer structures require no nanopatterning and thus scale more-easily to larger areas.

Active materials that have been recently investigated for tunable filters include correlated transition-metal oxides with phase transitions[19], germanium-antimony-tellurium (GST)[20], and gate-tunable graphene[21], all of which have complex refractive indices that can be tuned via one or more external stimuli, such as thermal biasing, electrical gating, incident light, and strain.[22,23,24] In particular, thin-film $VO_2$—the active medium used in this work—has been widely explored for tunable optical devices, especially in the mid- and far-infrared where it features a high refractive-index contrast, and has relatively low loss in one of its phases.[25] However, the integration of $VO_2$ into thin-film assemblies can be challenging because of the high temperatures required to obtain the correct crystallographic state (typically, >500°C [26,27,28]) and the significant dependence of its electrical and optical properties on the substrate choice and synthesis methods.[25,29]

In this paper, we demonstrated the fabrication of thin-film stacks comprising mid-infrared-transparent dielectric materials and $VO_2$, enabling the creation of tunable mid-infrared filters. In particular, we designed and fabricated an induced-transmission filter (ITF)—a dielectric thin-film filter that uses an ultrathin metal layer to remove unwanted sidebands around a desired passband—where the metal layer is replaced by $VO_2$. We designed filters with one or more tunable passbands, and experimentally demonstrated a filter that switches between a broadband transmission window in the long-wave infrared (LWIR, 8 – 12 µm) and a narrow passband centered around 8.8 µm.

**Design and Simulation**

Our design was developed based on the concept of induced-transmission filter (ITF) that was proposed by P. H. Berning and A. F. Turner in 1957.[30] Their original intent was to improve rejection of out-of-band transmission by introducing a thin lossy (metallic) film into a narrow-bandpass filter comprising dielectric thin films[30]. For example, Figure 1a is a schematic of a conventional single-cavity all-dielectric narrow-bandpass filter[31] that features a passband at free-space wavelength of $\lambda_0$, but is accompanied by unwanted side bands at both shorter and longer wavelengths. If an ultrathin (thickness $\ll \lambda_0$) layer of metal is inserted into the cavity at the position where the field magnitude is minimized (i.e., $|E|\sim 0$) at the desired wavelength, then the metallic film will have minimum influence on $E(\lambda_0)$ (i.e., $\varepsilon''|E|^2$ is small). On the other hand, the electric field at the other wavelengths can be greatly affected by this metallic film, resulting in efficient suppression of the side-band transmission. This configuration is an ITF with a single narrow passband at $\lambda_0$ (Figure 1b).

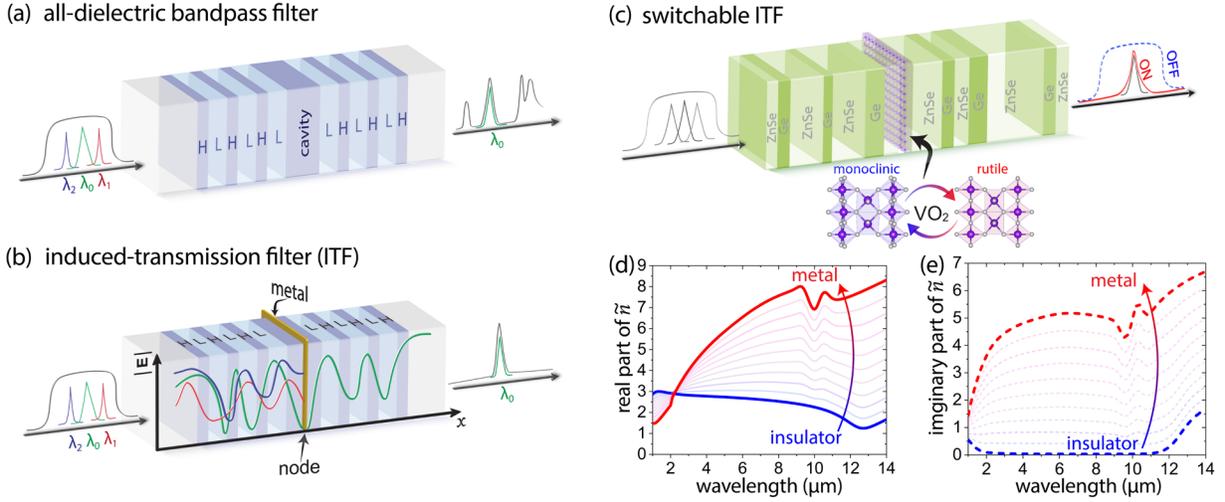

**Figure 1. (a)** A schematic of an all-dielectric single-cavity (Fabry-Perot-type) filter, with a high-index cavity sandwiched by alternating high-index (H) and low-index (L) dielectric films. This dielectric filter features a narrow transmission band but is accompanied by (usually unwanted) side bands at both shorter and longer wavelengths. **(b)** Schematic of an induced-transmission filter (ITF) in which an ultrathin metallic layer is introduced into the cavity at the position where the electric-field magnitude of the desired passband wavelength ($|E(\lambda_0)|$) is minimized, suppressing the unwanted sidebands. The schematics in (a, b) are rendered based on the design in ref.[31] **(c)** A schematic of our designed switchable ITF that features a broad transmission band (OFF state) when $VO_2$ is in the insulating phase and a narrow transmission band (ON state) when is $VO_2$ in the metallic phase. **(d, e)** The evolution of the complex refractive index of a ~100-nm $VO_2$ film across the insulator-to-metal transition (IMT)[25].

ITFs were invented and developed decades ago, but have been supplanted by advanced multilayer dielectric thin-film coatings.[32] Recent research into thin-film phase-transition materials present an opportunity to revisit this concept to enable a new class of tunable devices. Directly replacing the metal layer in a traditional ITF with a phase-transition material results in a modulation of the sideband transmission while leaving the primary passband at $\lambda_0$ essentially unchanged. In such switchable ITFs, those sidebands can be re-engineered to act as secondary passbands. Therefore, such a design would allow independent control of transmission in different passbands via tuning the phase of the active material, extending the potential uses of ITFs for tunable photonics.

Here, we use a $VO_2$ film as the switchable layer. $VO_2$ features a dramatic change in the complex refractive index in the infrared when the material undergoes a thermally driven insulator-to-metal transition (IMT) at ~70°C (Figure 1d, e).[25] As schematically shown in Figure 1c, once inserted into a bandpass filter, the insulator-phase $VO_2$ film is expected to have little effect on any transmission bands because of its low optical loss and deep-subwavelength thickness. In the metallic phase, the $VO_2$ film becomes very lossy and can suppress the transmission at all wavelengths except the wavelength at which the electric field has a node at the $VO_2$ position. We chose to demonstrate our design in the LWIR range of 8 to 12 μm, which is a window of atmospheric transparency, and is therefore a popular region for infrared imaging. We designed two switchable ITFs: a simpler design that can switch between double-passband and single-passband transmission states, and a more-complex design that switches between a narrow and a broad passband transmission state. The latter was then experimentally demonstrated.

To design a double-band transmission filter, we applied needle optimization methods[33,34] (implemented in OpenFilters[35]). The optimization started with a Bragg reflector structure that consists of alternating high- and low-index dielectric thin films. We targeted a transmission spectrum with two high-transparency windows at 9 and 11 μm, respectively. Since we were aiming for two transmission bands, we used two low-index materials (L), zinc sulfide (ZnS, $n$ = 2.2) and barium fluoride (BaF$_2$, $n$ = 1.42) and one high-index material (H), germanium (Ge, $n$ = 4). The optimized structure, with a refractive-index profile shown in Figure 2a, featured narrowband transmittance of ~1 at both target wavelengths.

We found that at a depth of 5.2 μm (Figure 2a), the magnitude of the electric field |$E$| featured a peak for incident wavelength λ = 9 μm and a node for incident wavelength λ = 11 μm, and thus this was the ideal position for insertion of thin-film VO$_2$ to selectively suppress transmittance at 9 μm. After we inserted a 50-nm VO$_2$ film at this position, the transmittance spectrum was largely unchanged for VO$_2$ in the insulating phase (OFF state). For VO$_2$ in the metallic phase (ON state), the transmittance at 9 μm was significantly suppressed, while the transmittance at 11 μm remained mostly unaffected (Figure 2b). The refractive index of VO$_2$ used in the simulation was extracted by ellipsometry analysis of a ~100-nm VO$_2$ film synthesized using the sol-gel method on a silicon (Si) wafer.[25]

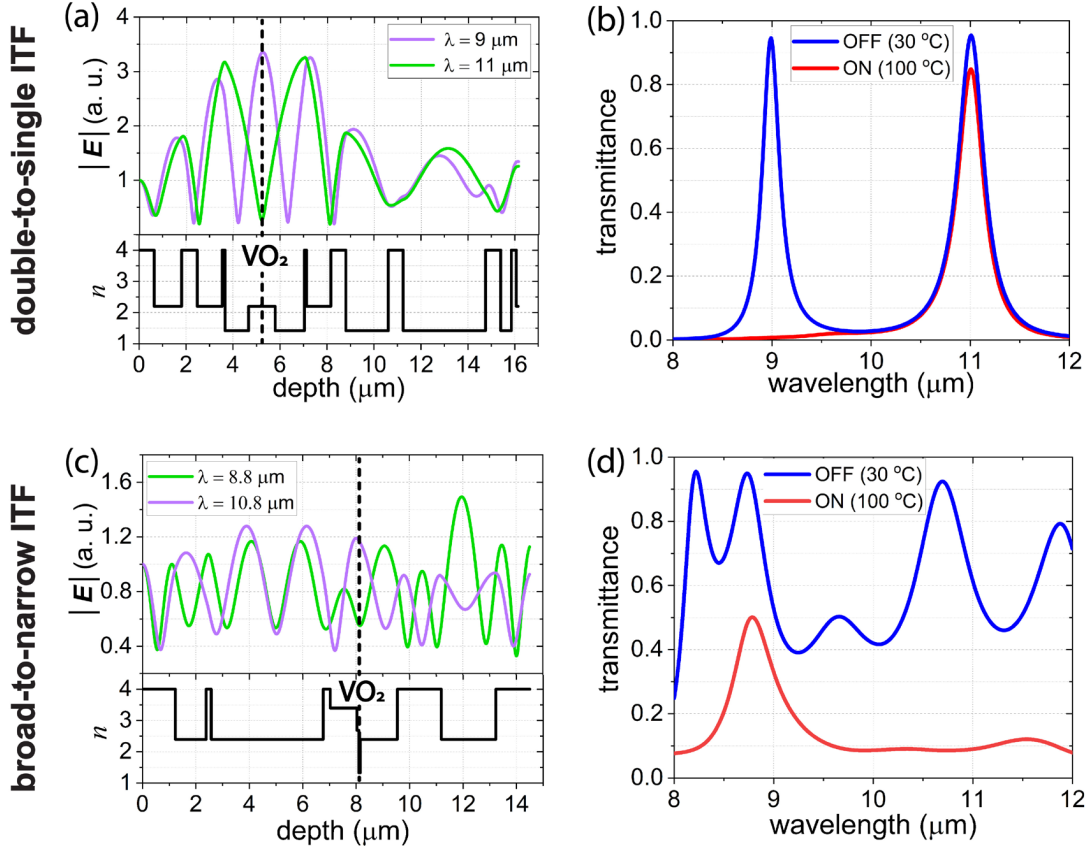

**Figure 2. (a)** Electric field distribution ($|E|$) in the dual-to-single bandpass ITF at wavelengths of 9 μm and 11 μm, when the $VO_2$ is in its insulating state. **(b)** Simulated OFF- and ON-state transmittance of the dual-to-single bandpass ITF. **(c)** Electric field distribution ($|E|$) in the broad-to-narrow bandpass filter at wavelengths of 8.8 μm and 10.8 μm, when the $VO_2$ is in its insulating state. **(d)** Simulated OFF- and ON-state transmittance of the broad-to-narrow bandpass ITF.

Then, via a similar procedure, we designed a broad-to-narrow bandpass ITF, with the target application of enhancing imaging using an infrared camera (see details in *Experiments*). We started with an all-dielectric stack with Ge ($n = 4.0$) as the high-index material and zinc selenide (ZnSe, $n = 2.4$) as the low-index material. In this case, we targeted high OFF-state transmittance ($T_{OFF}$) over a broad wavelength region (8 – 12 μm) and narrowband ON-state transmittance ($T_{ON}$) at ~9 μm. To minimize fabrication complexity, we limited the number of layers to be 12 and the total thickness to 14.5 μm. We set the thickness of $VO_2$ in the design to be 72 nm, to match a $VO_2$ film we previously synthesized on bulk Si substate using our sol-gel method, and imaged using cross-sectional SEM (*Supporting Information Section 1*). We also fixed the thickness of the Si substrate to be 1 μm to align with our fabricated device described in the next section. For the optimization, we set the figure of merit as the product of the OFF-state transmission ($T_{OFF}$) and the ratio between ON-state transmission ($T_{ON}$) within and outside of the passband at 8.8 μm:

$$FOM = T_{OFF}(8 - 12 \text{ μm}) \frac{T_{ON}^{max}(8.8 \text{ μm})}{T_{ON}(9.5 \text{ μm} - 12 \text{ μm})}.$$

The output structure (Figure 2c) featured an OFF-state LWIR transmission band consisting of five adjoining transmission bands with transmittance > 0.4 for all wavelengths and average transmittance > 0.6 from 8 to 12 μm (Figure 2d). In the ON state, the peak transmittance reaches 0.5 at 8.8 μm with out-of-band transmittance smaller than 0.1. Note that the FOM is maximized when the thickness of $VO_2$ is increased to 230 nm, however, a thicker $VO_2$ film also trades off the ON-state pass-band transmittance (see detailed discussion in *Supporting Information Section 2*). Therefore, one needs to find a good balance between the transmittance contrast and the absolute transmission magnitude.

**Experiments**

We fabricated the broad-to-narrow bandpass ITF based on the index profile shown in Figure 2c. The fabrication flow is described in Figure 3a. First, we used a 1-μm-thick Si (001) membrane supported by a 500-μm-thick frame (Norcada) as the substrate and oxidized it in a tube furnace at 850 °C for 5 hours in air to promote $VO_2$ adhesion. The suspended membrane was visibly wrinkled (Figure 3a-i). Our $VO_2$ films were produced by following a recipe based on the aqueous sol-gel method established by Hanlon *et al.*[36] A vanadium-oxide sol was prepared by heating vanadium pentoxide ($V_2O_5$) powder (99.99% purity, AlphaAesar) in a ceramic crucible at 1100 °C until molten, then slowly pouring into water and filtering out the precipitates. The $V_2O_5$ solution was then spin-coated onto the substrates to a thickness of ~70 nm, and baked at 150 °C to evaporate the solvent. After spinning, the substrate appeared multi-colored, with bands of red, green, and yellow across the substrate (Figure 3a-ii). The color variation corresponds directly to thickness variation and was caused by the wrinkled nature of the membrane substrate. The membrane was then placed inside a tube furnace in a reducing atmosphere of 5% $H_2$ in Ar and annealed for 550 °C for two hours to reduce the $V_2O_5$ film to $VO_2$ and promote crystal gain growth, forming high-quality $VO_2$ films.

The film uniformity was confirmed using scanning electron microscopy (SEM, Zeiss LEO 1530) and the phase transition of the as-synthesized $VO_2$ film was characterized using temperature-dependent microscopic Fourier-transform infrared (FTIR) spectroscopy (Hyperion 2000 coupled to Bruker Vertex 70) (*Supporting Information Section 3*). Then, we deposited the alternating Ge-ZnSe layers on the top and bottom sides of the $VO_2$-Si membrane (Figure 3a-iii and iv) using electron-beam evaporation (implemented by Materion Corp.).

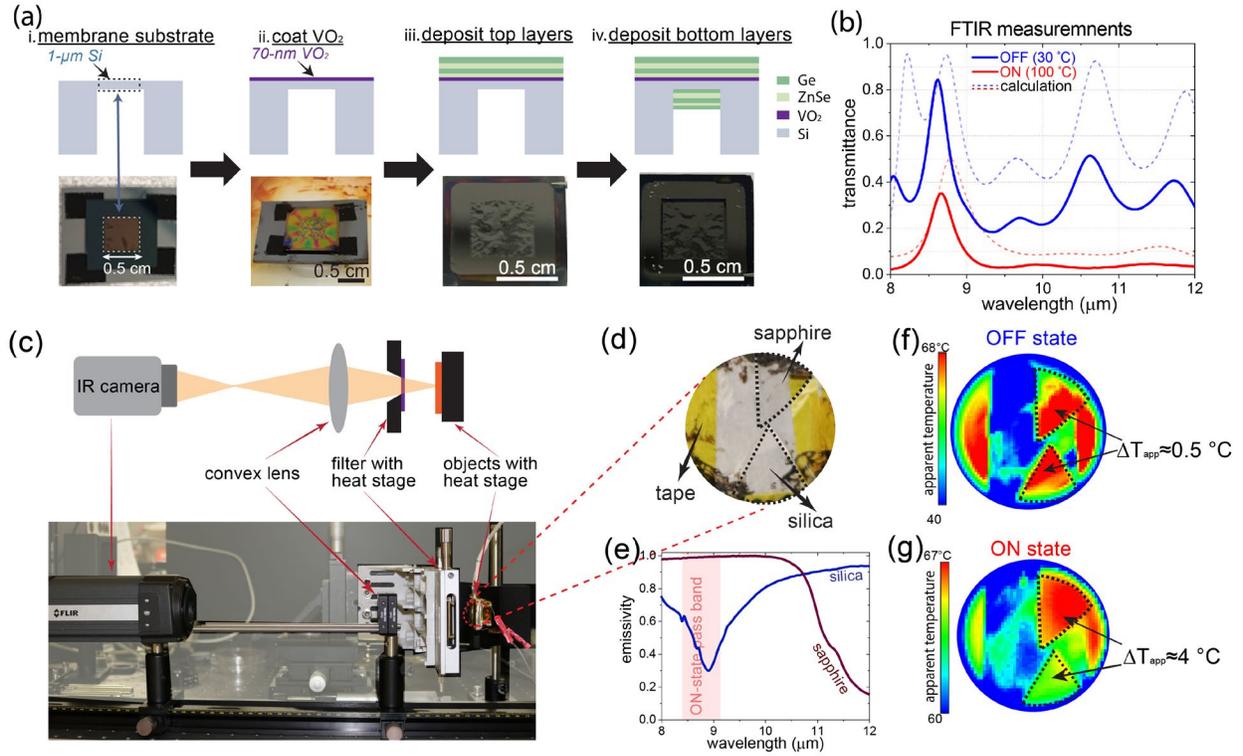

**Figure 3. (a)** Fabrication process. First, ~70-nm VO$_2$ was formed using the sol-gel method on the top side of a 1-μm thick Si membrane (5 × 5 mm) using sol-gel method. Then, the ZnSe/Ge layers were deposited on the top and bottom sides of the sample, with the target thicknesses according to our simulation shown in Figure 2c. **(b)** FTIR transmittance measurements of the fabricated ITF for both OFF (at 30 °C) and ON states (at 100 °C). The dashed lines are calculation results, same as those in Figure 2d. **(c)** Our infrared imaging setup: two test objects (sapphire and silica wafers) were heated to ~200 °C, and were imaged through the ITF, which was itself attached to a heat stage with a through-hole aperture. Then, a convex lens was used to collect the thermal emission and re-image the test objects to the object plane of an IR camera. **(d)** A photo of the sapphire and silica wafer pieces taken using a visible-light digital camera. **(e)** Emissivities of the sapphire and silica wafers at 200 °C, measured by our FTIR spectrometer using Kirchhoff's law (i.e., emissivity = absorptance = 1 – reflectance, for an opaque wafer)[37]. **(f, g)** Long-wavelength infrared (LWIR) images of the test objects with the filter in the OFF and ON states, respectively. When the filter was in the OFF state, the difference between the apparent temperatures of the sapphire and silica (ΔT$_{app}$) was only ~0.5 °C. When the filter was in the ON state, ΔT$_{app}$ increased to ~4 °C, making the two wafers substantially distinct.

The OFF- and ON-state transmittance of the fabricated ITF was characterized using our FTIR spectrometer with a heat stage (Linkam FTIR 600) used to control the temperature of the device. The measurements (solid curves in Figure 3b) are in qualitative agreement with our simulation (dashed curves in Figure 3b). At 30 °C (filter in the OFF state), the device features an average transmittance of 0.34 across 8 – 12 μm. At 100 °C (filter in the ON state), our filter has a narrow passband at λ = 8.7 μm, with peak transmittance of 0.35, accompanied by low out-of-band transmittance (< 0.03). The passband transmittance at 8.7 μm decreases by a factor of 2 as the device switches from the OFF state to the ON state, while the out-of-band transmittance decreases by an order of magnitude on average.

The ON-state passband of the fabricated filter is blue-shifted by 0.1 µm compared to the simulation, which can be attributed to differences in layer thicknesses between the model and the fabricated device. The thickness fluctuation also likely caused the lower transmittance of the fabricated device in both states compared to those of the simulation (*Supporting Information Section 4*). Moreover, there is possible presence of a small fraction of metallic $VO_{x<2}$ phases in the processed sol-gel film[38], resulting in higher optical loss that can also result in lower transmittance. Note also that small differences in the refractive indices are expected among $VO_2$ films, though they were synthesized via very similar conditions [25], which is another possibility for those discrepancies between the model and fabricated device.

Despite these discrepancies, our fabricated ITF still featured a good contrast in the transmittance bandwidth between the OFF and ON states, and can be used to enhance the imaging capability of a commercial LWIR camera when there is an imaging benefit to looking within a particular narrow wavelength range. As a demonstration, we incorporated our fabricated filter into an infrared imaging setup with a commercial FLIR A325sc camera (Figure 3c) which we used to image two test samples: pieces of sapphire and silica wafers. Sapphire and silica are nearly indistinguishable by visible-light cameras (Figure 3d), however, they have distinct emissivities in the LWIR region, especially at wavelengths within the ON-state passband of our fabricated ITF (Figure 3e).

The sapphire and silica wafers were taped to a heat stage, which was held at 200 °C. The thermally emitted light from the wafers went through the switchable ITF, which was attached to a separate heat stage with a through-aperture hole (Linkam FTIR 600). Then the transmitted power was collected using an infrared convex lens and re-imaged at the object plane of the LWIR camera. We set the camera emissivity to 0.95 for all the imaging (i.e., this is the emissivity the camera assumes when converting measured thermal-emission power into an apparent temperature), which is the default setting in the camera imaging software, and recorded images with the filter set to the OFF and ON states, respectively. When the filter in the OFF state (Figure 3f), the apparent temperature difference ($\Delta T_{app} = T_{sapphire} - T_{silica}$) between the sapphire and silica was very small (~0.5 °C), and thus the samples were almost indistinguishable, as they are in the visible. When the filter was switched to the ON state (90 °C), $\Delta T_{app}$ increased to 4 °C, thus enhancing the contrast between the two wafers, as shown in Figure 3g. Our imaging demonstration also indicates that our ITF design is somewhat tolerant to changes in the incident angle (*Supporting Information Section 5*).

Note that the apparent temperatures in the IR images were very different from the actual temperature of the objects (~200 °C), in part because we did not account for the true emissivities of the wafers, and in part because the optical components between the objects and camera (*i.e.*, the switchable ITF, the through-hole aperture in the heat stage, the convex lens) attenuated a significant amount of the emitted power from the objects. One can adjust the setting of the emissivity value in the camera software to make the apparent temperature close to the actual temperature, but this was not necessary for our demonstration.

**Discussion**

Our multilayer design geometry does not require nanopatterning, and thus this approach is more scalable and less expensive compared to another popular platform for tunable optical elements—active metasurfaces[39,40]. However, utilizing phase-change materials for tunable thin-film filters has not been broadly demonstrated[15,41], likely because it has been challenging to integrate the active materials into dielectric stacks. For example, tunable near-IR Bragg filters have been demonstrated by introducing thin-film $VO_2$ as a defect layer into an alternating $TiO_2/SiO_2$ optical stack by depositing all the layers on a bulk quartz substrate using conventional magnetron sputtering.[41] However, for the LWIR regime, very few transparent materials are convenient substrates for growth of $VO_2$[42,43]. Here, we demonstrated that a $VO_2$ film can be introduced into mid-IR-transparent stacks by first synthesizing $VO_2$ on a thin Si membrane and then depositing additional thin films on both sides. Therefore, our demonstration indicates the potential of using $VO_2$ for tunable thin-film filters over a broad range of wavelengths. We note that one could also use techniques such as the substrate transfer process[44,45] to fabricate the top and bottom thin-film coatings.

The concept of switchable ITFs is applicable for any desired wavelength range so long as there are suitable active materials and low- and high-index dielectrics that can be deposited to create the thin-film stack, such as the short-wave (0.9 to 2 µm), mid-wave (3 to 5 µm), and far (15 to 60 µm) infrared regimes. Besides $VO_2$, many phase-transition materials have been explored for different wavelengths of interest, including $Sb_2S_3$[46] and $1T-TaS_2$[47] for the visible, GST [23,15] for the near- and mid-infrared, and rare-earth perovskites[48] for the mid- and far-infrared. Moreover, we anticipate that more-sophisticated optimization methods[49,50,51] can be used to significantly improve device figures of merit compared to those of our proof-of-concept device. Finally, we note that electrically triggering the IMT of $VO_2$[52,53] can be used to replace the thermal biasing for better integration with other components.

**Conclusion**

In this paper, we generalized the concept of an induced-transmission filter (ITF)—an old filter design that used a thin metal layer to suppress extraneous transmission bands in dielectric thin-film filters—by replacing the metal film with a tunable phase-transition layer. We designed active ITFs that can switch from one passband to many using vanadium dioxide ($VO_2$), a phase-transition material with large optical contrast between its insulating and metal phases. Using a novel fabrication technique that combines sol-gel synthesis of $VO_2$ on a thin membrane with conventional electron-beam deposition of dielectric layers, we experimentally demonstrated an ITF that switches from a single narrow passband (at 8.8 µm) to a broadband transmission window (8 – 12 µm) within the long-wave infrared. Our active ITF is shown to improve infrared imaging contrast under certain conditions. Moreover, our design and fabrication methods provide a platform for realizing tunable thin-film filters for wavelengths of interest ranging from the visible to the far-infrared with the proper choice of the active materials, enabling applications such as multispectral imaging[54] and imaging through obscurants[55].

## Acknowledgement

This material is based on work supported by the Defense Advanced Research Projects Agency (DARPA) under Contract No. 140D6318C0013, and the paper was completed with support from the Office of Naval Research (ONR) under grant N00014-20-1-2297. The authors gratefully acknowledge use of facilities and instrumentation at the UW-Madison Wisconsin Centers for Nanoscale Technology (wcnt.wisc.edu) partially supported by the NSF through the University of Wisconsin Materials Research Science and Engineering Center (DMR-1720415).

# Supporting Information:

# Switchable induced-transmission filters enabled by vanadium dioxide

Chenghao Wan[1,2,†], David Woolf[3,†], Colin M. Hessel[3], Jad Salman[1], Yuzhe Xiao[1], Chunhui Yao[1], Albert Wright[3], Joel M. Hensley[3], Mikhail A. Kats[1,2,4*]

[1]Department of Electrical and Engineering, University of Wisconsin-Madison, Madison, WI 53706, USA
[2]Department of Materials Science and Engineering, University of Wisconsin-Madison, Madison, WI 53706, USA
[3]Physical Sciences, Inc., Andover, MA
[4]Department of Physics, University of Wisconsin-Madison, Madison, WI 53706, USA
[†]*These authors contributed equally to this work*
*Corresponding author: mkats@wisc.edu*

## Section 1. SEM imaging of the $VO_2$ film synthesized by sol-gel method

We synthesized a $VO_2$ film on a bulk undoped Si substrate using the sol-gel method, as elaborated in the main text. The film continuity and uniformity were confirmed by scanning electron microscopy (SEM) imaging of the top surface (Figure S1a). The film thickness was measured by SEM imaging of the cross section (Figure S1b). We also characterized the refractive indices of the film using infrared ellipsometry (ref. 25 in the main text). The film thickness and refractive indices used in the simulation were characterized based on this sample by assuming the properties of $VO_2$ on the Si membrane would be very similar with those of $VO_2$ on the bulk Si substrate because they were prepared via the same procedures.

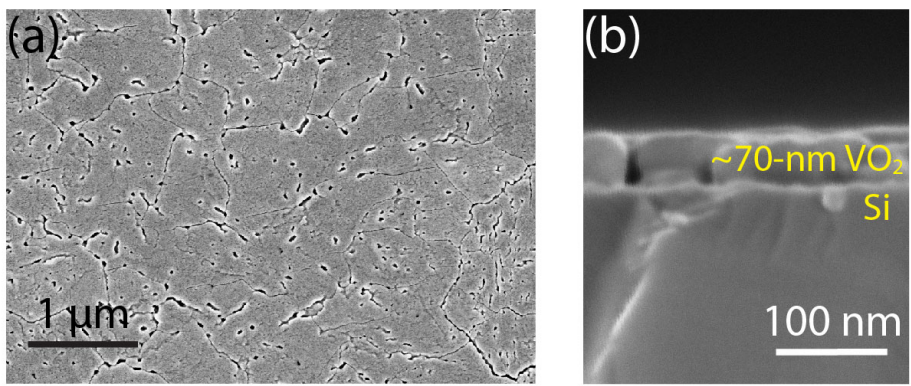

**Figure S1**. SEM imaging of the **(a)** top surface and **(b)** cross section of the $VO_2$ deposited on bulk Si substate using the sol-gel method.

## Section 2. The dependence of ITF performance on the thickness of $VO_2$:

In our design of the broad-to-narrow bandpass ITF, the thickness of $VO_2$ was not optimized to maximize the figure of merit (FOM) because we assumed it has little influence on the pass-band transmittance due to its deep-subwavelength thickness. As shown in Figure S2a, we found that the OFF-state spectrum is almost independent on the VO$_2$ thickness ranging from 50 nm to 340 nm, while the ON-state transmittance is gradually reduced as the thickness increases, resulting a maximum FOM when the thickness of VO$_2$ reaches 230 nm (Figure S2b). However, the ON-state pass-band transmittance is reduced to ~0.2 for this maximum FOM, which could make it challenging to image through the filter (e.g., the proof-of-concept IR imaging demonstration in Figure 3f and g in the main text). Therefore, one may need to balance between the transmittance contrast and the absolute transmission magnitude for practical uses.

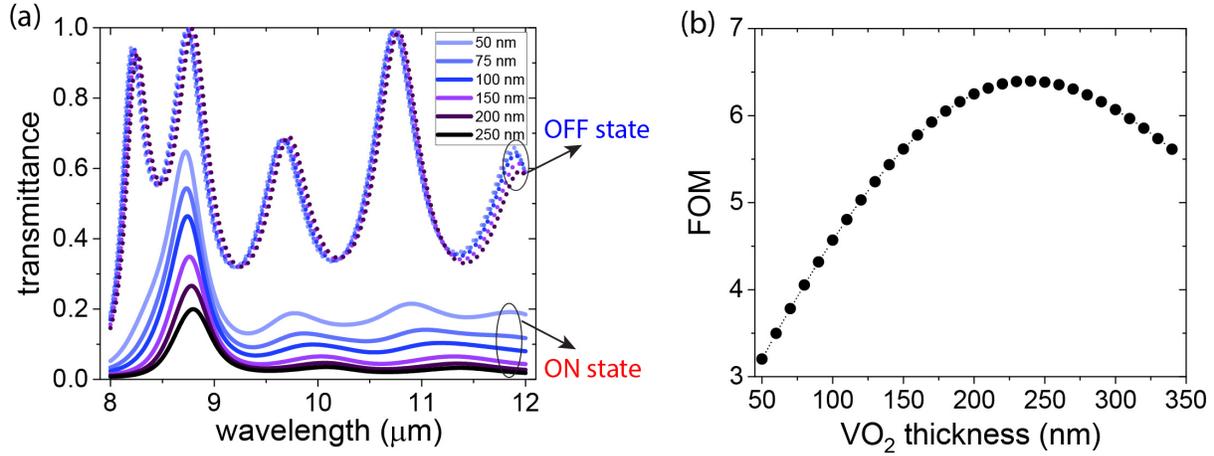

**Figure S2. Dependence of our broad-to-narrow bandpass ITF performance on the thickness of VO$_2$. (a)** Calculated OFF- and ON-state spectra for the thickness of VO$_2$ set to be 50, 75, 100, 150, 200, and 250 nm. **(b)** Calculated FOM for VO$_2$ thickness ranging from 50 to 340 nm.

### Section 3. Temperature-dependent FTIR transmittance measurements

We performed temperature-dependent transmittance measurements on both the as-synthesized VO$_2$/Si membrane (Figure S3a, b) and the final broad-to-narrow ITF (Figure S3c, d), using an infrared microscope (Bruker Hyperion 2000) that is connected to a Fourier-transform infrared (FTIR) spectrometer (Bruker Vertex 70). All spectra were collected at temperatures between 30 °C and 90 °C (first heating, then cooling) with steps of 2 °C. For a thermally driven device, one should avoid biasing the filter at temperatures in the thermal hysteresis region shown in Figure S3b, d. It is also worth noting that the VO$_2$/Si membrane behaves as a decent tunable Fabry-Perot (FP) interferometer (a.k.a., etalon) with a broadband transmittance window at central wavelength of ~7 μm at room temperature (i.e., VO$_2$ in the insulating phase) and low transmittance (~0.05) at 90 °C (VO$_2$, in the metallic phase), which may be useful on its own.

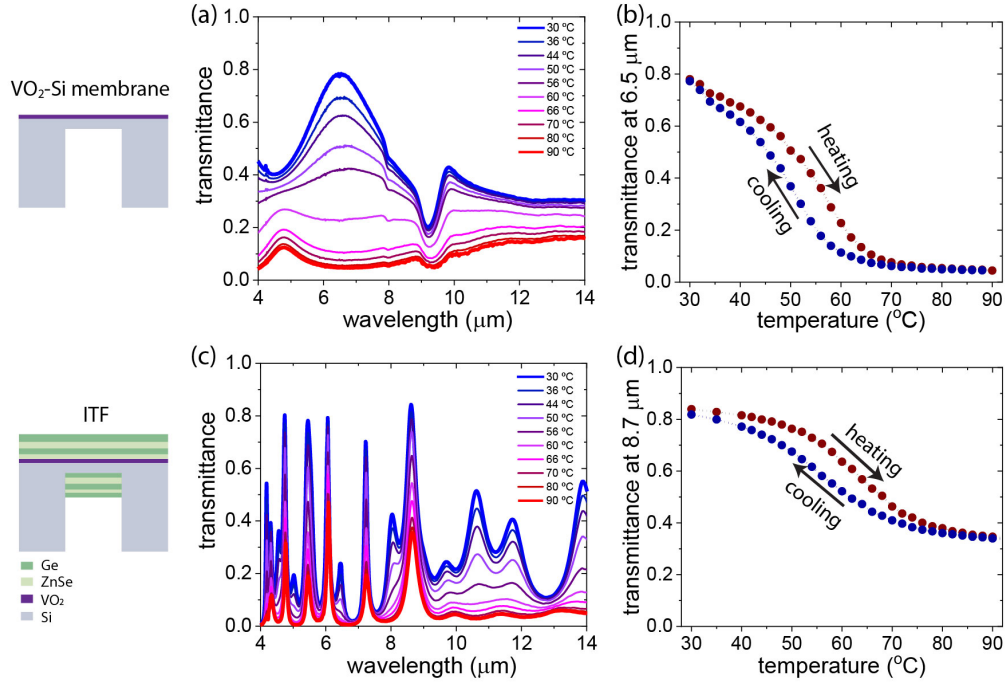

**Figure S3**. Temperature-dependent transmittance of the **(a, b)** post-synthesized VO$_2$/Si membrane (i.e., the step ii in Figure 3a) and **(c, d)** the fabricated ITF (i.e., step iv in Figure 3a). (a) and (c) are measurements taken during a heating process from 30 °C to 90 °C, with a ramping rate of 1 °C/min. (b) and (d) are the evolution of the transmittance peak across the IMT, taken in a cycle of heating and cooling, with a ramping rate of 1 °C/min.

## Section 4. Influence of the thickness fluctuation

We estimated the tolerance of our designed broad-to-narrow ITF with respect to thickness variations in all of the layers, which are inevitable during the fabrication. We applied three sets of variations: +5% in thickness for each layer, -5% in thickness for each layer, and randomly +5% or -5% in thickness for each layer. As shown in Figure S4, these thickness variations can cause either a shift of the transmission peaks or a reduction of the transmission magnitudes, or both, and thus are likely the primary reason for the transmittance differences between the simulated and fabricated ITF shown in Figure 3b. Note that the results in Figure S4 were calculated using our own transfer-matrix codes at a different time than the calculations using OpenFilters in Figure 3b, and there may have been minor differences in the refractive indices assumed. Therefore, we observed some minor differences in the transmittance spectra between Figure S4 and Figure 3b.

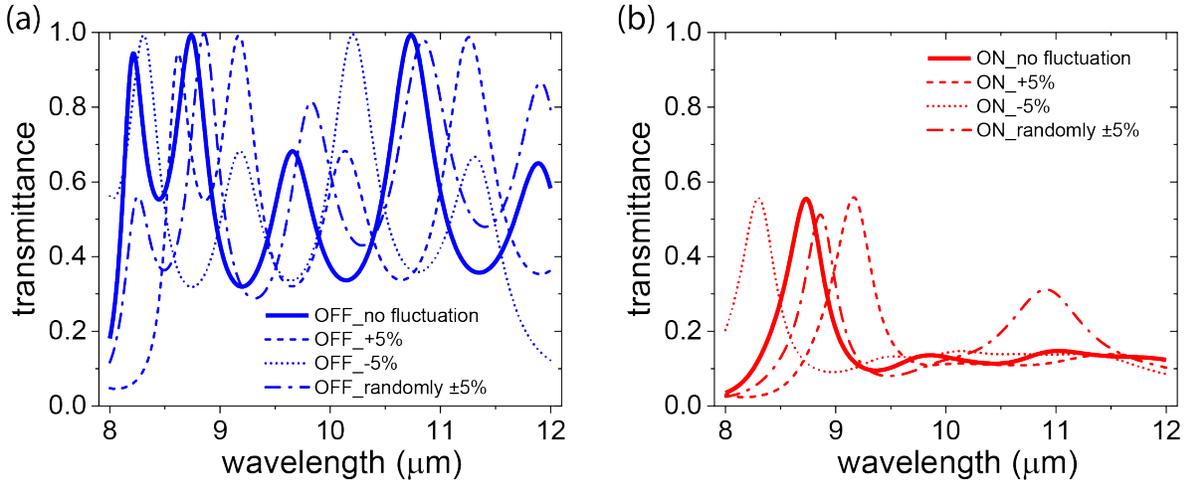

**Figure S4**. Influence of thickness fluctuations on both **(a)** OFF-state and **(b)** ON-state transmittance spectrum of the broad-to-narrow ITF in Figure 2c, d. The four sets of curves assume no fluctuations in the thickness of the layers (solid), a uniform increase in thickness of all layers by +5% (dashed line) and -5% (dotted line), and random thickness fluctuations from layer to layer by +5% or -5% (dashed-dotted line).

## Section 5. Incident-angle dependence of the broad-to-narrow ITF

Incident-angle-dependent transmittance spectra of our designed broad-to-narrow ITF were calculated using the transfer-matrix method for both p- and s-polarizations as shown in Figure S5 a and b, respectively. For both OFF and ON states, the transmittance spectra do not change much for incident angles smaller than 20 degrees, which enabled us to acquire thermal images of the silica and sapphire wafers through the filter.

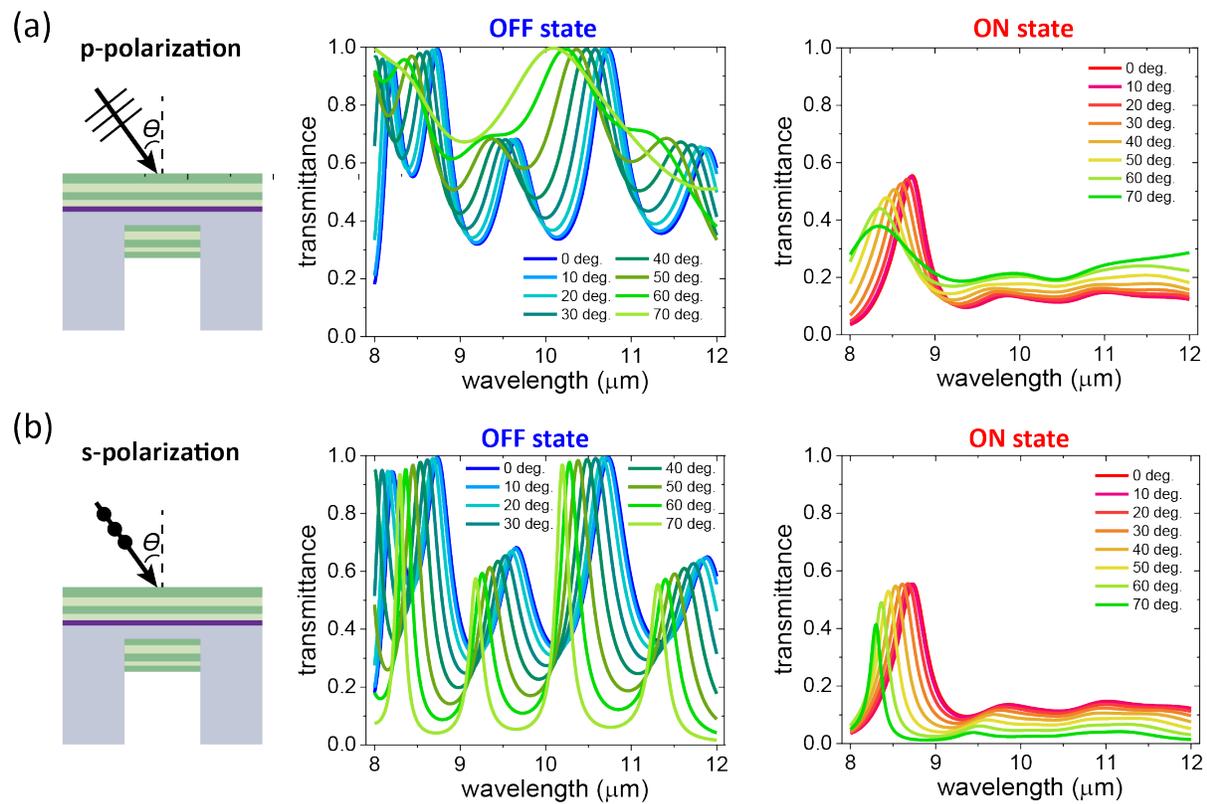

**Figure S5.** Calculated incident-angle-dependent transmittance spectra of the broad-to-narrow ITF for both **(a)** p-polarization and **(b)** s-polarization.